# SOUND-BASED DRONE FAULT CLASSIFICATION USING MULTITASK LEARNING

Wonjun Yi, Jung-Woo Choi and Jae-Woo Lee
*Korea Advanced Institute of Science and Technology (KAIST), Daejeon, Republic of Korea*
email: lasscap@kaist.ac.kr

The drone has been used for various purposes, including military applications, aerial photography, and pesticide spraying. However, the drone is vulnerable to external disturbances, and malfunction in propellers and motors can easily occur. To improve the safety of drone operations, one should detect the mechanical faults of drones in real-time. This paper proposes a sound-based deep neural network (DNN) fault classifier and drone sound dataset. The dataset was constructed by collecting the operating sounds of drones from microphones mounted on three different drones in an anechoic chamber. The dataset includes various operating conditions of drones, such as flight directions (front, back, right, left, clockwise, counterclockwise) and faults on propellers and motors. The drone sounds were then mixed with noises recorded in five different spots on the university campus, with a signal-to-noise ratio (SNR) varying from 10 dB to 15 dB. Using the acquired dataset, we train a DNN classifier, 1DCNN-ResNet, that classifies the types of mechanical faults and their locations from short-time input waveforms. We employ multitask learning (MTL) and incorporate the direction classification task as an auxiliary task to make the classifier learn more general audio features. The test over unseen data reveals that the proposed multitask model can successfully classify faults in drones and outperforms single-task models even with less training data.
Keywords: drone, fault classification, direction classification, multitask learning

## 1. Introduction

Recently, drones have been used in many different fields. In manufacturing, for example, drones are crucial in automating facility management and condition monitoring. Drones are also actively utilized for military, disaster sites, and agricultural applications. Nevertheless, drones are prone to mechanical faults, and slight damage to their sensors, propellers, or motors can make them unable to accomplish given tasks. Thus, fault detection and diagnosis in the early stages are essential for the safe and reliable operation of drones.

Many studies have been done on the mechanical fault diagnosis of drones by processing data from vision sensors [1]. However, faults in drones mainly occur in fast-rotating propellers or motors, so vision-based fault diagnosis requires images of high resolution, resulting in increased costs. Instead of vision sensors, vibration sensors and microphones are also being utilized for fault diagnosis of drones [2]–[5]. Microphones can especially handle a wide frequency range, which is beneficial for detecting high-frequency faults occurring by fast-rotating propellers or motors.

However, research on fault diagnosis of drones using sounds suffers from the lack of real-world datasets. Some prior works utilized microphones placed outside of drones [3], [4], fixed drones on a tripod [4], or drones flying indoor spaces with reflective walls nearby [3], [5]. In real-world applications, however, drones typically operate at high altitudes outdoors, and their faults can occur after the take-off. Consequently, microphones must be mounted on drones for real-time fault diagnosis during the flight. Also, unrealistic environmental interferences such as wall reflections should be suppressed. Lastly, operating drones exhibit various movements, so the dataset needs to include data acquired with multiple flight conditions of drones.

In this work, we propose a dataset of drone operating sounds that can be utilized for training a deep neural network (DNN) model to diagnose drone faults. To overcome the limitation of previous studies, we constructed the proposed datasets as follows; (1) we collected a large audio dataset using three heterogeneous quadcopters with various faults and maneuvering conditions. A total of six maneuvering directions (forward, backward, right, left, clockwise, counterclockwise) and nine operating statuses (normal, fault in one of four propellers, fault in one of four motors) were considered. (2) Operating sounds of drones were recorded in an anechoic chamber to reduce sound reflections from walls. (3) To synthesize realistic drone sounds contaminated by noises, background noises were separately recorded in outdoor venues.

In addition to the dataset, we introduce a DNN model that can robustly diagnose drone faults irrespective of their maneuvering conditions. Since rotations of motor-propeller pairs are different depending on flight maneuvers, the fault diagnosis must be made regarding the flight condition. We demonstrate that a DNN classifier subjected to solving multiple tasks related to the flight condition can learn the relation between the maneuvering directions and the status of drones, which allows for the successful diagnosis of drone status.

## 2. Dataset

### 2.1 Materials

Figure 1 displays three types of drones used for experiments, referred to as type A, B, and C, from left to right, respectively. Drone type A is Holy Stone HS720, B is MJX Bugs 12 EIS, and C is ZLRC SG906 pro2. They have different frame sizes, weights, specifications of motors, and lengths of propellers.

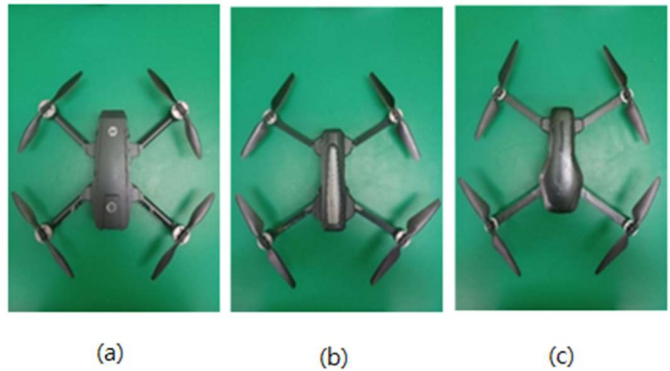

Figure 1: Three drone types used for the experiment.
(a) Type A (Holy Stone HS720), (b) Type B (MJX Bugs 12 EIS), (c) Type C (ZLRC SG960 pro2)

We made faulty drones by breaking down parts of normal drones in two different ways: cutting the propeller and denting the motor cap. In the case of the *propeller cut*, we cut about 10% of a single propeller (Fig. 2(a)). Because only one propeller out of four was broken, the drone cannot balance and tends to generate abnormal vibrations during rotation. For *dented motor cap*, we crushed the motor cap supporting the rotor and protecting the coil by applying strong force using a vice (Fig. 2(b)). The crushed cap adds more friction and prevents the smooth rotation of a motor since the cap.

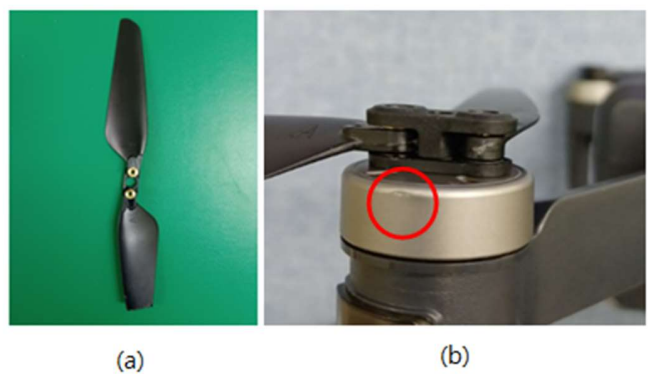

Figure 2: Faults of drone type B. (a) propeller cut, (b) dented motor cap (red circle indicates dented part)

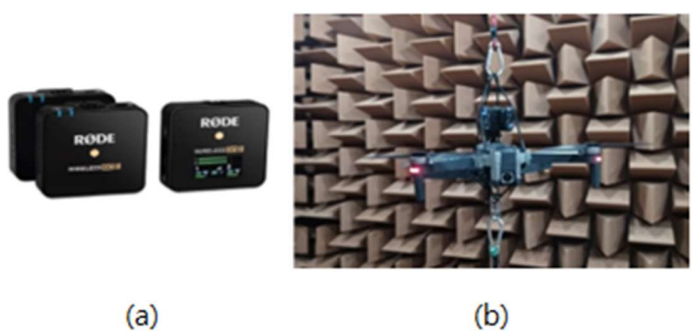

Figure 3: (a): Røde Wireless Go2 microphones (transmitter, receiver), (b): recording sounds of drone type B

## 2.2 Recording

We used the dual-channel wireless microphone (RØDE Wireless Go2) to record the operating sounds of drones. Each microphone was attached to the top side of the drone body, and its sensitivity was adjusted to avoid the clipping of high-level operating sounds from drones. The operating sounds of drones were recorded in the anechoic chamber, as shown in Fig. 3(b), to minimize sound reflections from walls. During the recording, drones were loosely tied by two elastic ropes of 1.5 m and 1 m length connected to the ceiling and floor, respectively. Rotatable rings were used to combine the drone and rope such that the interference of the ropes on the drone movement could be minimized. Background noise signals were collected using the same microphone at five different places on the university campus (Fig. 4).

## 2.3 Dataset

Operating sounds of drones and background noises were recorded with a sampling rate of 48 kHz but were down sampled to 16 kHz because most of the frequency components appeared below 8 kHz. After down-sampling, we separated each sound recording into segments of 0.5 seconds. We then mixed the operating sounds of drones and background noises with different SNRs varying from 10 dB to 15 dB. SNR is expected to be high because the microphone is directly attached to drones emitting loud operation sounds.

We constructed a dataset of 54k noisy drone signals for each drone type. Each data was labeled with nine drone status labels (normal, propeller cut on each propeller, the dented motor cap for each motor) and six maneuvering direction labels (forward, backward, right, left, clockwise, and counterclockwise). Both the drone status label and maneuvering direction label were given by one-hot vectors. For each drone type, the dataset was randomly shuffled and split into the train, valid, and test datasets in a 6:2:2 ratio, respectively. The dataset can be downloaded via Zenodo (https://zenodo.org/record/7779574#.ZCOvfXZBwQ8).

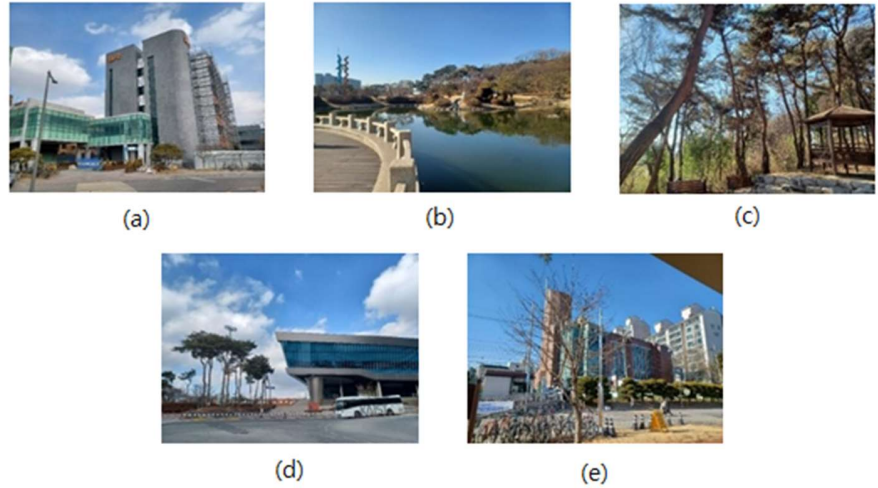

Figure 4: Five different spots on the university campus chosen for background noise recording: (a) construction site, (b) pond, (c) hill, (d) sports complex, and (e) gate.

## 3. Sound-based drone fault classification using multitask learning

We designed and trained a sound-based DNN fault classifier using the constructed dataset. The main objective of the DNN classifier is to predict the drone status label for fault diagnosis. The drone operating sound varies with respect to the rotation speed of the propellers and motors on the four arms of quadcopters, which depend on the maneuvering of drones as well as the drone status. To reflect the data variations according to maneuvering directions, we configured the auxiliary task of classifying maneuvering direction labels. The DNN model was designed as a multitask model with a shared feature extractor whose output is handled by two heads predicting the status label and the maneuvering direction label, respectively. The overall architecture of the designed fault classifier is schematized in Fig. 5.

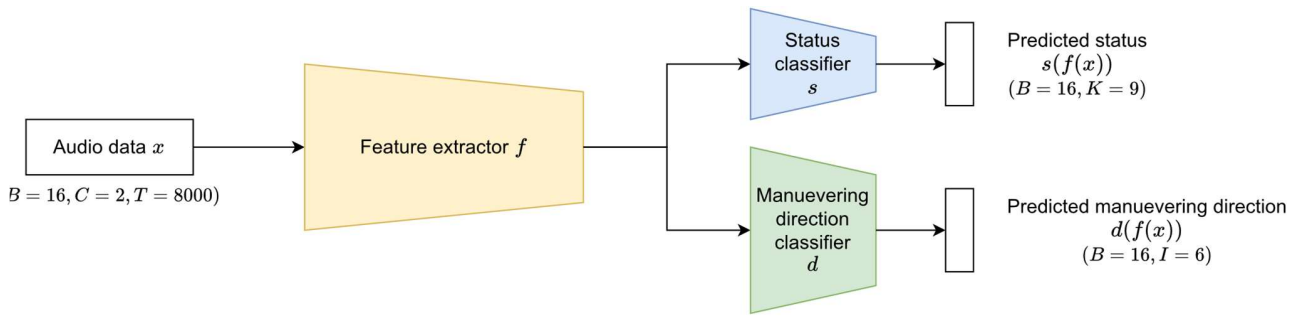

Figure 5: Fault classifier trained by multitask learning

In Fig. 5, the audio data $x \in \mathbb{R}^{B \times C \times T}$ with batch size $B$, channel size $C$, and the number of time samples $T$ are fed into the feature extractor $f$ and transformed into the feature vector $f(x)$. The feature vector becomes the shared input of the status classifier $s$ and maneuvering direction classifier $d$. It is then converted to the softmax probability of predicted status $s(f(x)) \in \mathbb{R}^{B \times K}$ and predicted maneuvering direction $d(f(x)) \in \mathbb{R}^{B \times I}$, respectively, for $K$ status labels and $I$ maneuvering directions. Using the predicted softmax probabilities and the target probability of status $y_s$ and maneuvering direction $y_d$, we define the drone status loss $\mathcal{L}_s$ and maneuvering direction loss $\mathcal{L}_d$ as

$$\mathcal{L}_s = E[H(y_s, s(f(x)))], \quad \mathcal{L}_d = E[H(y_d, d(f(x)))], \tag{1}$$

where $E[\cdot]$ denotes the sample mean over batch indices and $H(\cdot)$ represents categorical cross entropy (CCE). The weighted sum of two losses gives the total loss $\mathcal{L}_{tot}$ defined as

$$\mathcal{L}_{tot} = \frac{1}{\sigma_s^2}\mathcal{L}_s + \frac{1}{\sigma_d^2}\mathcal{L}_d + \ln(1+\sigma_s^2) + \ln(1+\sigma_d^2), \tag{2}$$

where observation noise parameters $\sigma_s^2$ and $\sigma_d^2$ are learnable weights employed for balancing losses $\mathcal{L}_s$ and $\mathcal{L}_d$ [6], respectively. The last two terms of Eq. (2) are logarithm regularization terms blocking the divergence of observation noise parameters.

## 4. Experiment

We compared the status classification performance of the multitask learning (MTL) and single-task learning (STL) models. The STL model excludes the projection head for classifying the maneuvering direction and is trained only to classify the status of drones. In the first experiment, we examined the classification performance for each drone type to check the generalization ability of the proposed model across drones of different specifications. Next, the performance change with respect to the size of the training dataset was investigated. This experiment was for inspecting the influence of small datasets regarding the difficulties in collecting a large drone dataset in practical applications. The investigation was conducted with datasets of various sizes, i.e., using 100%, 50%, 25%, and 10% of the total dataset.

### 4.1 Network architecture

In the proposed DNN architecture, the input is first processed by 1D convolutional layers that extract spectral information from the measured waveform. The spectra of drone operating signals deliver essential information, e.g., the harmonics produced by the propellers appearing from 250 Hz to 300 Hz and the harmonics from motors varying from 2 kHz to 3 kHz. The 1D convolutional layers act as learnable band-pass filters, which are initialized as 64 gammatone filters [7]. Max-pooling is applied to the absolute value of the output signal from the 1D convolutional layer to reduce the data size in the following feature extractor. As the feature extractor, we employed 1DCNN-ResNet with 1D convolution filters for its relatively low computational complexity and a small number of parameters. To enhance the classification performance of vanilla 1DCNN-ResNet, we added an average pooling layer on the residual connection when the number of output channels differs from the number of input channels in the corresponding residual block [8], and squeeze and excitation (SE) blocks on the main path for every residual block [9]. The output of the last residual block is averaged in the channel direction and fed to the status and maneuvering direction classifiers consisting of a single linear layer and softmax activation.

### 4.2 Train and validation

We trained the model for 100 epochs using the Adam optimizer and learning-rate scheduler with an initial learning rate of $5 \times 10^{-4}$. The batch size was 16, and the stochastic weight averaging [10] was applied at 80 epochs. We validated the model's performance using the validation dataset for each epoch and stored the best model for the test. As the performance metric, we used the Macro F1 score given by $F1_{Macro} = \frac{1}{N}\sum_{n=1}^{N} F1_n$ for $N$ class labels.

## 5. Result

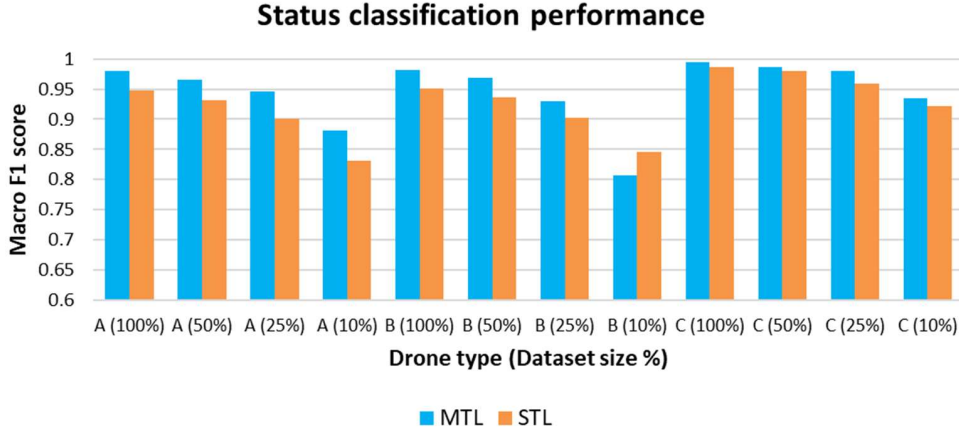

Figure 6: Status classification performance of MTL and STL models.
Abscissa indicates the drone type and dataset size, and the ordinate displays the F1 score.

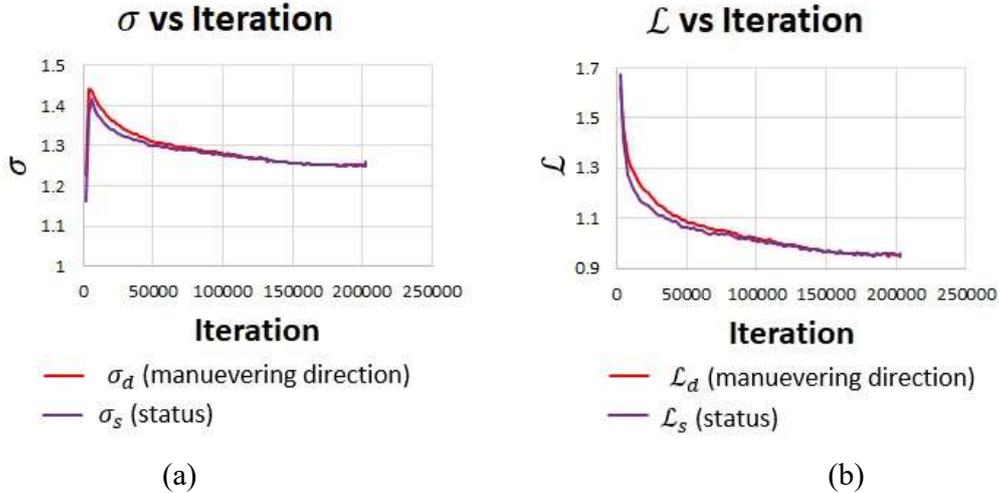

Figure 7: (a) Observation noise parameters ($\sigma_d$, $\sigma_s$) and (b) losses ($\mathcal{L}_d$, $\mathcal{L}_s$) with respect to iterations. Results are for the MTL model and drone type A trained by 100% dataset. The red lines are the results of maneuvering direction classification and the purple lines indicate the results of status classification.

The comparison of MTL and STL models in terms of status classification performance is presented in Fig. 6. Except for model B (10%), MTL outperforms STL for every drone type and dataset size, which demonstrates that the features extracted for finding out maneuvering directions also strengthens the ability to diagnose the drone status. The MTL model can successfully classify the drone status with the macro F1 score higher than 95%, even when only 50% of the train dataset is used. In Fig. 7, the maneuvering direction loss is slightly greater than the status loss, and the observation noise parameter for maneuvering direction ($\sigma_d$) also has a larger value than that for the drone status. This result implies that the higher maneuvering direction loss is weighted by the higher observation noise parameter to secure the balance with the drone status loss. When macro F1 scores are compared (Fig. 8), the maneuvering direction classification seems to be a harder task than the drone status classification. From the comparison of Fig. 6 and Fig. 9, we can also see that the maneuvering direction classification is not as accurate as the drone status classification. Nevertheless, the training of the MTL model with this difficult auxiliary task brings a significant increase in the macro F1 score over the STL model. These results stress that the use of drone maneuvering information in the DNN classifier training not only improves the fault diagnosis performance but also enables the model training with less data.

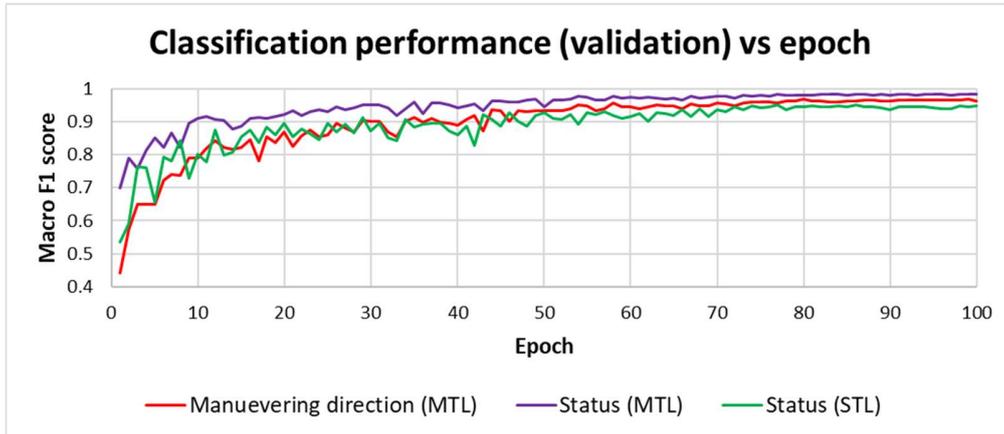

Figure 8: F1 score measured in every epoch during validation. The red and purple lines are the F1 scores of the MTL model calculated for the maneuvering direction classification and drone status classification tasks, respectively. For comparison, the F1 score of the STL model for the drone status classification task is presented as a green line (all models were trained on 100% dataset).

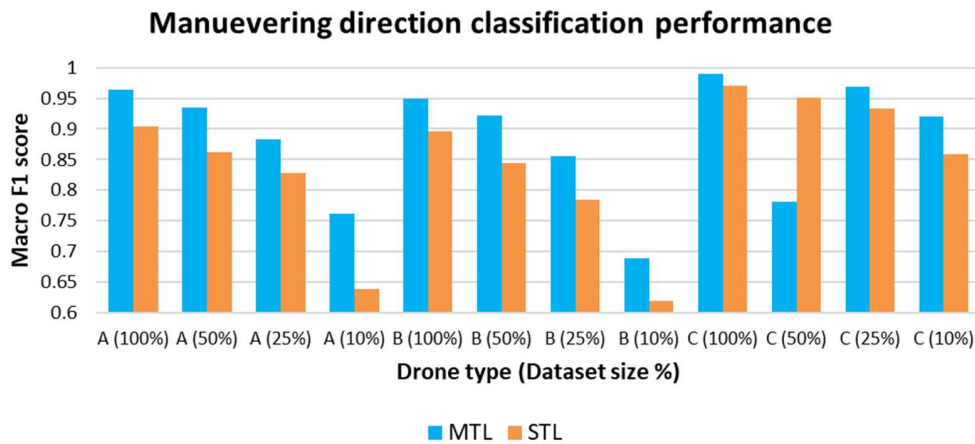

Figure 9: Classification performance for maneuvering directions. Abscissa indicates drone type and dataset size, and ordinate presents the F1 score.

## 6. Conclusion

We constructed the operating sound dataset of three quadcopter drones labelled with drone statuses and maneuvering directions. The sound data from drones of nine different statuses (normal, propeller cuts, and dented motor caps) were measured for various drone maneuvering directions (forward, backward, right, left, clockwise, and counterclockwise). We recorded drone sounds in the anechoic chamber to simulate acoustically dry conditions and mixed them with background noises measured in outdoor environments. Along with the dataset, we also proposed a DNN-based fault classifier that was designed to accomplish multitask objectives of classifying drone statuses and maneuvering directions. The experiments with the proposed dataset and DNN model showed that the multitask model can achieve higher status classification performance than the single-task model for different types of quadcopters and train dataset sizes. It was also found that maneuvering direction classification is a more complicated task than drone status classification, and the multitask model gains the ability to classify faulty drones better while learning to solve the more difficult task. The DNN model can be improved further by utilizing spatial information using multichannel audio data. However, the large drone sound dataset proposed in this work can be used for versatile tasks related to the anomaly detection and condition monitoring of drones.

## ACKNOWLEDGEMENTS

This work was supported by the BK21 Four program through the National Research Foundation (NRF) funded by the Ministry of Education of Korea.